\documentclass[prbr,twocolumn,showpacs,superscriptaddress]{revtex4-1}
\usepackage{graphicx}
\usepackage{bbm}
\usepackage{bbold}
\usepackage{amssymb}

\begin{document}

\title{Surface instability in nodal noncentrosymmetric superconductors}

\author{Carsten Timm}
\email{carsten.timm@tu-dresden.de}
\affiliation{Institute of Theoretical Physics, Technische Universit\"at
Dresden, 01062 Dresden, Germany}

\author{Stefan Rex}
\affiliation{Department of Physics, Norwegian University of Science and
Technology, 7491 Trondheim, Norway}

\author{P. M. R. Brydon}
\email{pbrydon@umd.edu}
\affiliation{Condensed Matter Theory Center and Joint Quantum
Institute, University of Maryland, College
Park, USA 20742}

\date{April 8, 2015}

\begin{abstract}
We study the stability of topologically protected zero-energy flat
bands at the surface of nodal noncentrosymmetric
superconductors, accounting for the alteration of the gap near the
surface. Within a selfconsistent mean-field theory, we show that the
flat bands survive in a broad temperature range below the bulk
transition temperature. There is a second transition at a lower
temperature, however, below which the system spontaneously
breaks time-reversal
symmetry. The surface bands are shifted away from zero energy and become
weakly dispersive. Simultaneously, a spin polarization and an equilibrium
charge current develop in the surface region.
\end{abstract}


\maketitle

\textit{Introduction}.\ The topological properties of gapless electronic systems
have recently attracted much attention
\cite{ScR11,SBT12,ZhW13,MCS13,YPZ14,ScB15}. An important example are
time-re\-ver\-sal-sym\-me\-tric noncentrosymmetric superconductors (NCSs)
\cite{Sat06,Ber10,Tanaka2010,STY11,BST11,ScR11,SBT12}, which are characterized
by strong
antisymmetric spin-orbit coupling (SOC) and a parity-mixed pairing state
\cite{ncsbook}. Many NCSs display evidence of gaps with line nodes
\cite{Izawa2005,Yuan2006,Nishiyama2007,Mukuda2008,Bonalde2009,Eguchi2013}. This
is exciting, as the line nodes of NCSs with dominant triplet pairing are
topologically nontrivial defects in momentum space
\cite{ScR11,SBT12,ZhW13,MCS13}. Zero-energy flat bands of Majorana fermions are
predicted to appear within the projections of these nodal lines onto the surface
Brillouin zone (BZ). Such flat bands have clear experimental signatures such as
sharp zero-bias peaks in tunneling spectra \cite{BST11,SBT12}, equilibrium
currents parallel to the interface between the NCS and a ferromagnet
\cite{BTS13,STB13}, and characteristic quasiparticle interference
patterns~\cite{HQS13}.

The topological properties of NCSs and consequently the protection of the
surface states are controlled by the superconducting gaps, which arise from
interactions. Properly accounting for these interactions may qualitatively
alter the surface physics. For example, a surface tends to suppress some gap
components and enhance others \cite{MaS95,FRS97,Sig98,ZFT99,HWS00,KaT00,BGB13}.
This may change the conclusions of the aforementioned studies
\cite{Tanaka2010,STY11,BST11,ScR11,SBT12,ZhW13,MCS13,YPZ14,ScB15},
which imposed unrealistic uniform gaps. Flat bands with their high density of
states are particularly prone to instabilities. Indeed, the zero-energy flat
bands at the $(110)$ surface of \textit{d}-wave superconductors with
time-reversal symmetry (TRS) \cite{TaK95} are predicted to be unstable
towards a time-reversal-symmetry-breaking (TRSB) state
\cite{MaS95,FRS97,Sig98,ZFT99,HWS00,KaT00,BGB13,PL14}.
This has been supported by some tunneling and transport
experiments \cite{Cov97,KD99,GGF13} but was not seen in others
\cite{WYG98,BFQ02,KKP04,CSD05,WKK08}.
\textit{d}-wave superconductors are however qualitatively different from NCSs
in that the zero-energy flat bands are degenerate in the first case but
nondegenerate in the second.

In this paper, we study the stability of the surface zero-energy flat bands of
nodal NCSs by performing selfconsistent mean-field (MF) calculations in real
space for a slab of finite thickness. For concreteness, we consider a model with
point group $C_{4v}$, which is realized for $\mathrm{CePt}_3\mathrm{Si}$
\cite{BHM04}, $\mathrm{CeRhSi}_3$ \cite{KIS05}, and  $\mathrm{CeIrSi}_3$
\cite{SOS06}. We show that an instability to a TRSB state can occur and study
its signatures.

\textit{Model and mean-field theory}.\ We start from a tight-binding Hamiltonian
for an NCS with $C_{4v}$ point group, $H=H_0+H_\mathrm{int}$. The noninteracting
part is
\begin{eqnarray}
H_0 &=& -\mu \sum_j c_j^\dagger c_j - t \sum_{\langle ij\rangle}
  (c_i^\dagger c_j + c_j^\dagger c_i) \nonumber \\
&& {}+ i\lambda \sum_{\langle ij\rangle} (\hat\mathbf{z} \times
  \hat\mathbf{e}_{ij}) \cdot \left(
  c_i^\dagger\, \frac{\mbox{\boldmath$\sigma$}}{2}\, c_j
  - c_j^\dagger\, \frac{\mbox{\boldmath$\sigma$}}{2}\, c_i \right) ,
\label{H0.2}
\end{eqnarray}
with the chemical potential $\mu$, the nearest-neighbor hopping amplitude $t$,
and the Rashba SOC strength $\lambda$. The SOC term breaks inversion symmetry.
The annihilation operator $c_j = (c_{j,\uparrow},c_{j,\downarrow})^T$ is a
two-component spinor, {\boldmath$\sigma$} is the vector of Pauli matrices, and
$\hat\mathbf{e}_{ij}$ is the unit vector pointing from site $j$ to site $i$ of a
simple cubic lattice. Attractive interactions at the same site and between
nearest neighbors in the $xy$ plane are described by
\begin{equation}
H_\mathrm{int} = -U_s \sum_j c^\dagger_{j\uparrow} c^\dagger_{j\downarrow}
  c_{j\downarrow} c_{j\uparrow}
  - U_t\!\! \sum_{\langle ij\rangle\perp\hat\mathbf{z}} \sum_{\sigma\sigma'}
  c^\dagger_{i\sigma} c^\dagger_{j\sigma'} c_{j\sigma'} c_{i\sigma} .
\end{equation}
The interaction is decoupled in the pairing channel. We define the singlet and
triplet order parameters
$\Delta^s_j \equiv (U_s/2)\, \langle c_j^T i\sigma^y c_j\rangle$ and
$\mbox{\boldmath$\Delta$}^t_{ij} \equiv iU_t\, \langle c_j^T i\sigma^y
\mbox{\boldmath$\sigma$} c_i\rangle$,
respectively, where the site indices $i$, $j$ in
$\mbox{\boldmath$\Delta$}^t_{ij}$ are restricted to nearest-neighbor sites in
the $xy$ plane. The triplet vector order parameter is taken to be parallel
to the effective SOC field,
$\mbox{\boldmath$\Delta$}^t_{ij} = \Delta^t_{ij}\, \hat\mathbf{z}\times
\hat\mathbf{e}_{ij}$.
This choice avoids the triplet-pair-breaking effect of the SOC, and is
therefore e\-ner\-ge\-ti\-cal\-ly favorable in the bulk~\cite{FAK04}.

\begin{figure}
\includegraphics[scale=0.2]{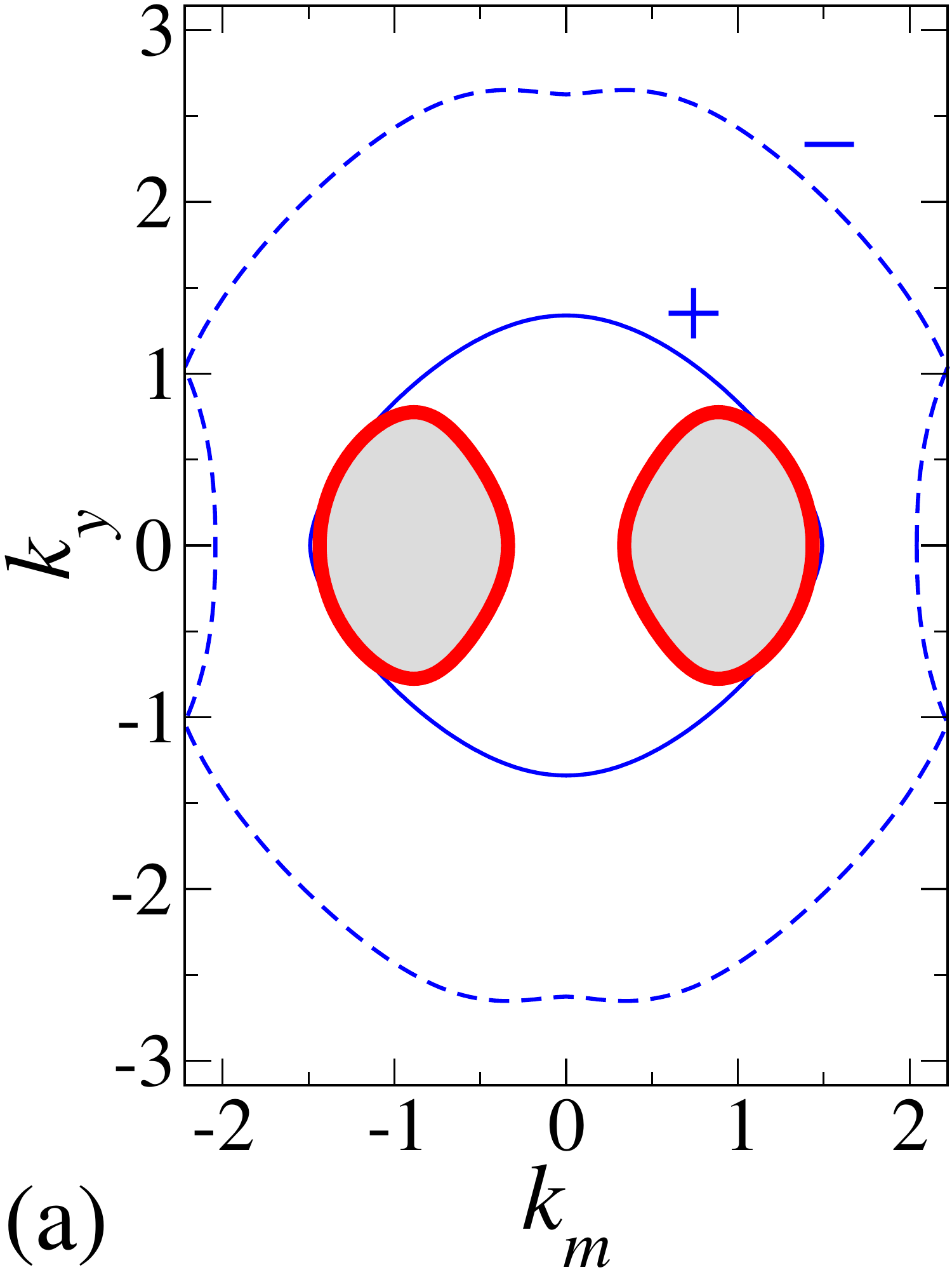}\hspace{1em}%
\includegraphics[scale=0.2]{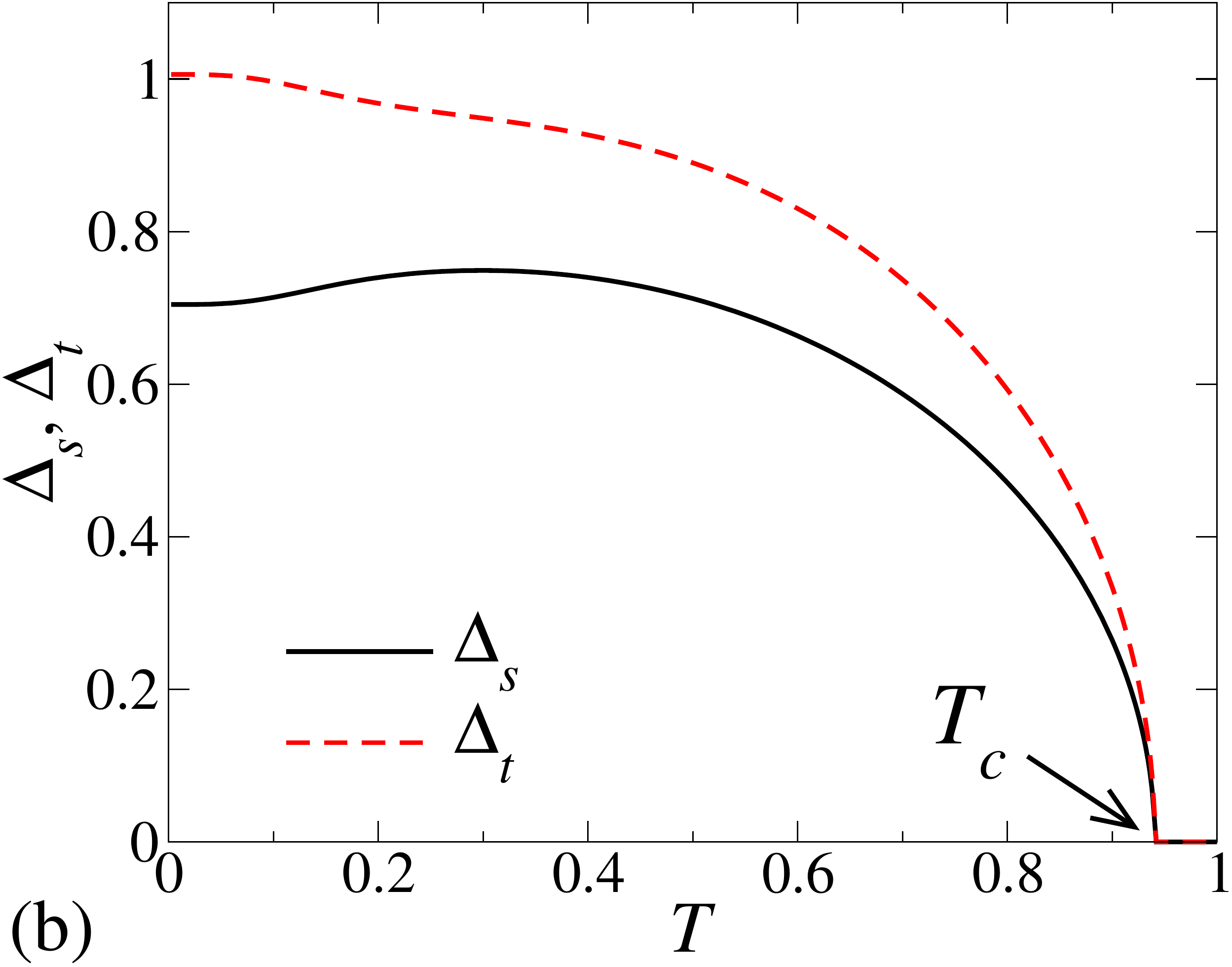}
\caption{(Color online) (a) Projection onto the (101) plane of the
positive-helicity Fermi surface (thin solid line), the negative-helicity Fermi
surface (dashed line), and the superconducting gap nodes on the former (heavy
solid lines), for the bulk NCS. The gray areas denote the zero-energy flat bands
predicted to exist at (101) surfaces under the assumption of uniform gaps
\cite{ScR11,SBT12}. The plot is restricted to momenta in the (101) surface BZ,
where $k_m = (k_x-k_z)/\sqrt{2}$. The parameters are $t=1$, $\lambda=1.5$,
$\mu=-3$, $U_s=5.0$, $U_t=5.4$, and $T=0.0025$. (b) Mean-field gaps $\Delta_s$
(solid black) and $\Delta_t$ (dashed red) as functions of temperature $T$.}
\label{fig.Fermisurf}
\end{figure}

We first consider the MF solution for an extended system, assuming spatially
uniform gaps $\Delta^s_{j}=\Delta_s$ and $\Delta^t_{ij} = \Delta_t$. Details of
the calculation are given in Sec.~{I} of the Supplemental Material~\cite{suppl}.
We find that the singlet and triplet gaps have the same phase, which can be set
to zero, so that TRS is preserved. SOC splits the bands and thus also the Fermi
surface according to the helicity of states \cite{SBT12}. Since the triplet
order parameter is parallel to the SOC, pairing only occurs between states with
the same helicity.
We determine interaction strengths $U_s$, $U_t$ that lead to flat zero-energy
surface bands under the assumption of uniform gaps. The resulting surface states
have been studied in detail in Refs.\ \cite{BST11,SBT12,BST15}.
This is realized for the parameters $t=1$ (hence, $t$ is our unit of energy),
$\lambda=1.5$, $\mu=-3$, $U_s=5.0$, $U_t=5.4$ at the temperature $T=0.0025$
(setting $k_B=1$), giving bulk MF gaps $\Delta_s = 0.704$ and $\Delta_t =
1.006$. We consequently find a gap with line nodes on the (smaller)
positive-helicity Fermi surface, but a full gap on the (larger)
negative-helicity Fermi surface~\cite{endnote.lambda}. Figure
\ref{fig.Fermisurf}(a) shows the projection of the two Fermi surfaces and the
nodal lines onto the (101) plane. The topological argument from Refs.\
\cite{ScR11,SBT12} predicts that a (101) surface hosts flat zero-energy bands
within the region bounded by the projected nodal lines. In addition, there is an
arc of zero-energy states connecting the two regions with flat bands
\cite{BST11,SBT12,YPZ14}. Figure \ref{fig.Fermisurf}(b) shows the bulk gaps
$\Delta_s$ and $\Delta_t$ as functions of temperature.

\begin{figure}
\includegraphics[width=0.85\columnwidth]{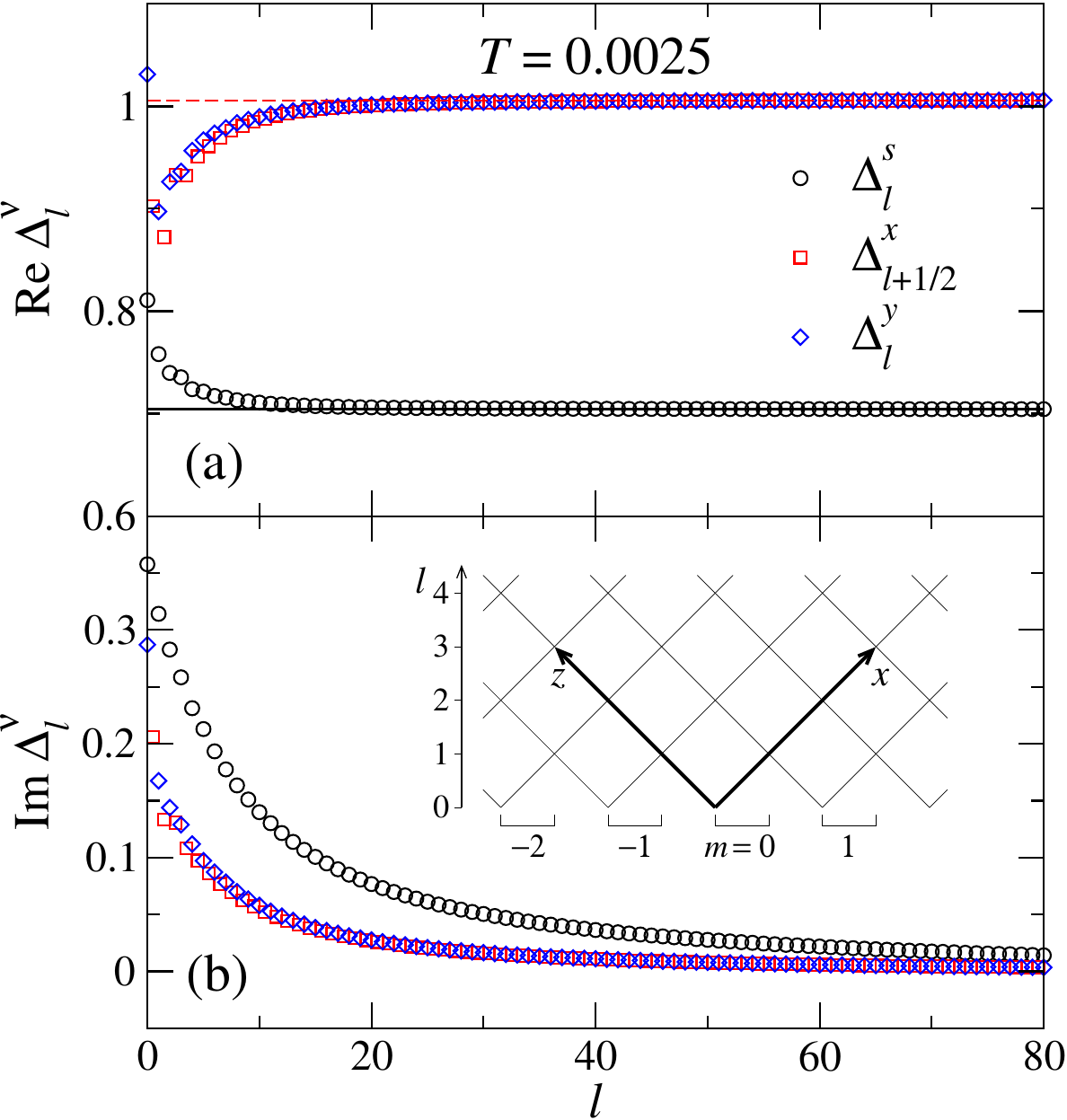}
\caption{(Color online) Selfconsistent gaps $\Delta^s_l$, $\Delta^x_{l+1/2}$,
$\Delta^y_l$ for a slab of thickness $W=300$ and parameters as in Fig.\
\ref{fig.Fermisurf}. (a) and (b) show the real and imaginary parts,
respectively. The lines denote the bulk gaps $\Delta_s$ (solid black) and
$\Delta_t$ (dashed red). Inset: Sketch of the bottom ($l=0$) surface of a (101)
slab, showing the new coordinates $l$ and $m$. The $y$ axis points into the
plane of the drawing.}
\label{fig.selfcgaps}
\end{figure}

We next turn to the MF solution for a slab of thickness $W$ with (101)
surfaces. We introduce new coordinates
$x = m + (l+l\,\mathrm{mod}\,2)/2$ and
$z = -m + (l-l\,\mathrm{mod}\,2)/2$,
where $m$ is parallel to the surfaces and $l=0,\ldots,W-1$ is orthogonal to
them. The geometry of one surface and our coordinate system are depicted in the
inset of Fig.\ \ref{fig.selfcgaps}. Since translational symmetry in the normal
direction is broken, the gaps depend on $l$. We define
\begin{eqnarray}
\frac{U_s}{2}\, \langle c_j^T i\sigma^y c_j\rangle &\equiv& \Delta^s_l ,\\
iU_t\, \langle c_j^T i\sigma^y
  \mbox{\boldmath$\sigma$} c_i\rangle &\equiv&
  \left\{\begin{array}{ll}
    \Delta^x_{l+1/2}\, \hat\mathbf{z}\times \hat\mathbf{e}_{ij} &
    \mbox{for $x$ bonds,} \\[1ex]
    \Delta^y_l\: \hat\mathbf{z}\times \hat\mathbf{e}_{ij} &
    \mbox{for $y$ bonds,}
  \end{array}\right.
\end{eqnarray}
where the subscript $l$ denotes the (identical) $l$ coordinate of sites $i$ and
$j$, while $l+1/2$ in $\Delta^x_{l+1/2}$ is the mean of the $l$ coordinates of
sites $i$ and $j$. We Fourier transform in the directions parallel to the slab,
introducing the two-dimensional momentum vector $\mathbf{k} = (k_m,k_y)$ in the
surface BZ, defined by $-\pi < k_y \leq \pi$ and $-\pi/\sqrt{2} < k_m \equiv
(k_x-k_z)/\sqrt{2} \leq \pi/\sqrt{2}$. The MF calculations are performed for a
slab of thickness $W=300$, using the same parameters as for the bulk
calculation. Further details are presented in Sec.\ {II} of the Supplemental
Material~\cite{suppl}.

\begin{figure}
\includegraphics[width=0.85\columnwidth]{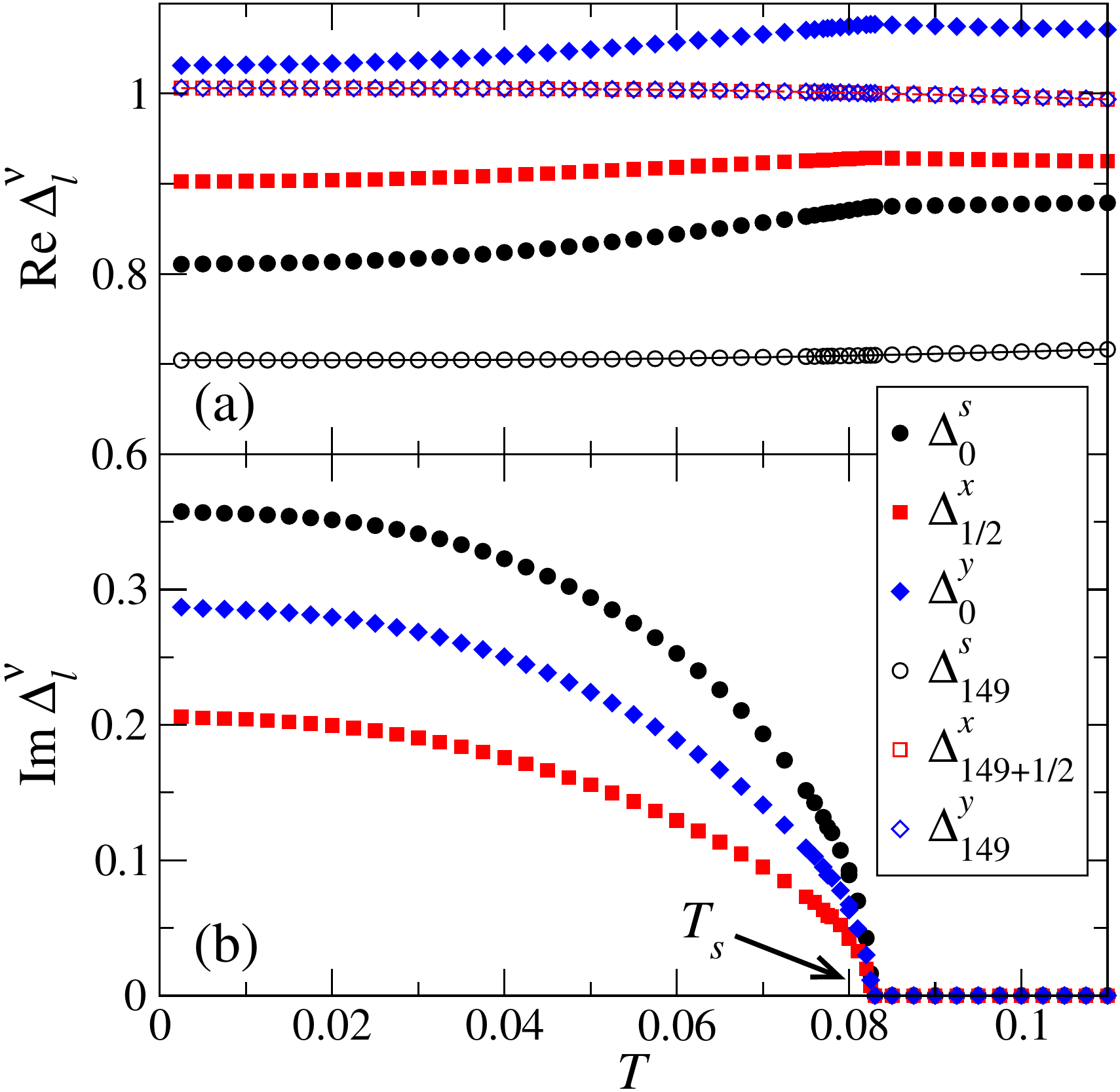}
\caption{(Color online) Selfconsistent gaps $\Delta^s_l$, $\Delta^x_{l+1/2}$,
$\Delta^y_l$ for the surface layer ($l=0$, filled symbols) and at the slab
center ($l=W/2-1$, open symbols) as functions of temperature. The thickness is
$W=300$, the parameters are as in Fig.\ \ref{fig.Fermisurf}. (a) and (b)
show the real and imaginary parts, respectively. The imaginary parts for
$l=W/2-1$ would be indistinguishable from zero and are omitted. The lines in
panel (a) denote the bulk gaps $\Delta_s$ (solid black) and $\Delta_t$ (dashed
red) from Fig.~\ref{fig.Fermisurf}(b).}
\label{fig.gapsT}
\end{figure}

\textit{Spontaneous breaking of TRS}.\ Our central results are summarized in
Figs.\ \ref{fig.selfcgaps} and \ref{fig.gapsT}: at sufficiently low
temperatures, the singlet and triplet gaps develop imaginary components close to
the surface, spontaneously breaking TRS. This solution is degenerate with a
state with complex-conjugated gaps. In the limit $W\to\infty$, the two surfaces
are decoupled and there are hence four degenerate TRSB solutions, differing in
the signs of the imaginary parts of the gaps close to the surfaces.

The spatial variation of the gaps near the surface in the TRSB phase is shown in
Fig.~\ref{fig.selfcgaps}. While both the sing\-let and trip\-let gaps develop
imaginary components near the surface, the real parts of the singlet and triplet
gaps are enhanced above and suppressed below their bulk values, respectively.
The suppression of the triplet gaps originates from the pair-breaking effect of
the surface, which in turn enhances the singlet gap to compensate for the lost
condensation energy. The reversal of the suppression of the triplet gaps in the
outermost layer can be understood similarly: since one of the triplet amplitudes
is missing at the surface, the others are enhanced.

The gaps converge to their bulk values as we move away from the surface; the
gaps at the center of the slab are within $0.01$\% of their bulk values. Note
that the deviation of the imaginary parts from their bulk value (of zero) has a
much longer range than that of the real parts. Indeed, close to the center of
the slab, we find that $\text{Im}\, \Delta^\nu_l \propto (l - W/2)$, see Fig.\
\ref{fig.selfcgaps}(b). We have checked that the proportionality constant
decreases more rapidly than $W^{-1/2}$ with $W$ so that the gradient energy
vanishes for $W\to\infty$. We attribute the slow spatial decay to the
enhancement of length scales close to the bulk quantum phase transition to a
nodeless singlet-dominated state. This transition can be reached by increasing
$U_s$ and decreasing $U_t$ by only $0.067$ (not shown).

The evolution of the TRSB state with temperature is shown in
Fig.~\ref{fig.gapsT}, where we plot the gaps $\Delta^s_l$, $\Delta^x_{l+1/2}$,
and $\Delta^y_l$ in the surface layer and at the slab center. Upon increasing
the temperature, the gaps in the surface layer show a second-order transition,
at which the imaginary parts vanish and TRS is restored. This occurs at a
temperature of $T_s \approx 0.083$, well below the bulk superconducting
transition temperature $T_c \approx 0.942$.

\begin{figure}
\includegraphics[width=\columnwidth]{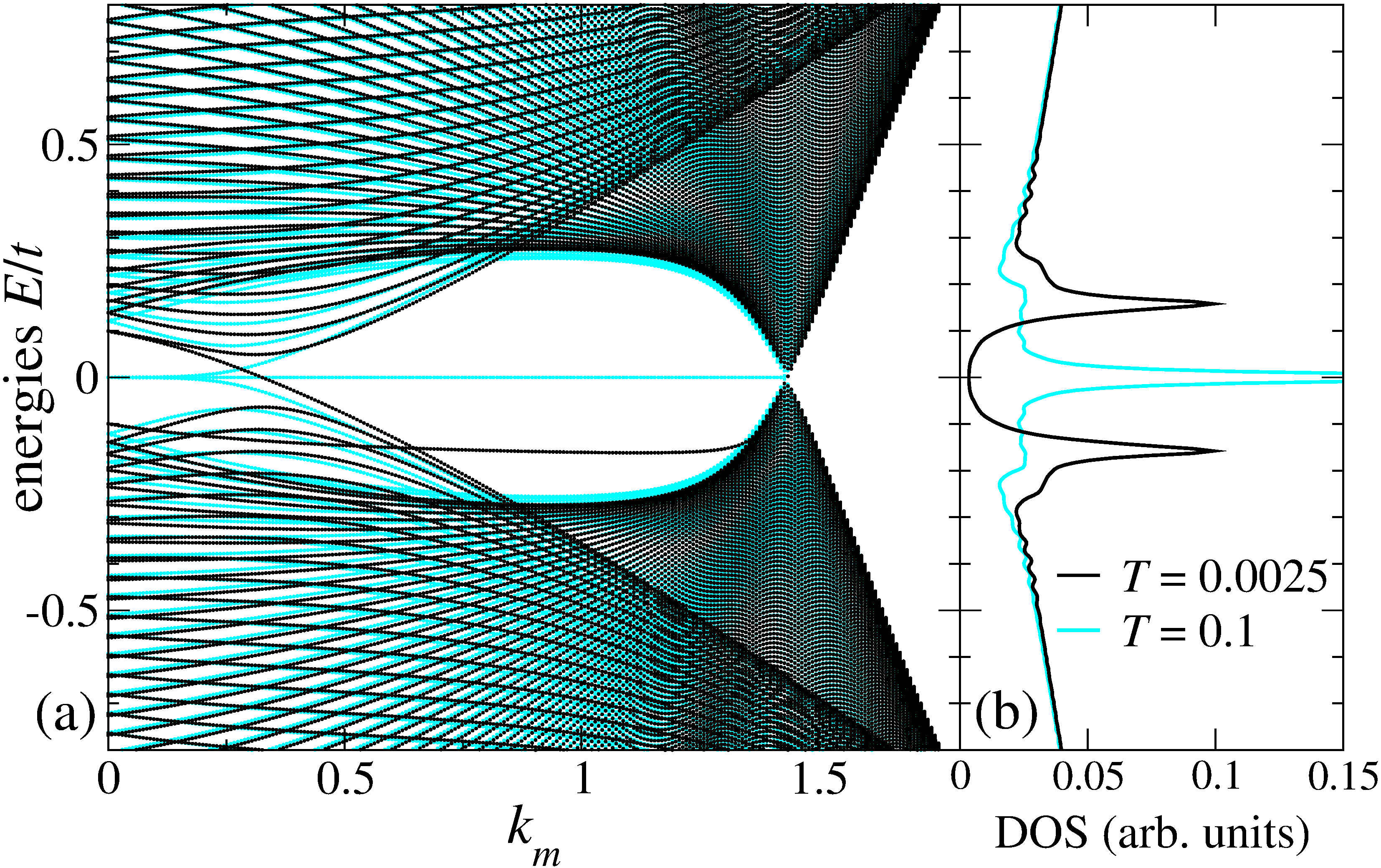}
\caption{(Color online) (a) Dispersion for a cut through the surface BZ at
$k_y=0$, for $W=300$ and the same parameters as in Fig.\ \ref{fig.Fermisurf}.
The black points refer to $T=0.0025\ll T_s$ in the TRSB state, whereas the cyan
(light gray) points in the background refer to $T=0.1>T_s$ with restored TRS.
The dispersion is odd in $k_m$, only points for $k_m\ge 0$ are shown. (b)
Surface DOS in the $l=0$ layer at the same temperatures. An artificial
broadening of $\eta=0.01$ was used.}
\label{fig.disp}
\end{figure}


\textit{Dispersion and density of states}.\ In Fig.\ \ref{fig.disp}(a) we plot
the dispersion for a cut through the surface BZ at $k_y=0$ at temperatures below
and above $T_s$. For $T>T_s$, the zero-energy flat band predicted in Refs.\
\cite{BST11,SBT12} is clearly visible for $0.5 \lesssim k_m \lesssim 1.5$; the
zero-energy states at $k_m \lesssim 0.5$ form an arc connecting the projections
of the nodal rings \cite{BST11,SBT12}. The TRSB for $T<T_s$ removes the
topological protection of the zero-energy flat bands of the TRS state, which are
consequently pushed away from zero energy, with a low-temperature energy shift
on the order of $T_s$. Since the shift is weakly momentum dependent, the band
obtains a nonzero velocity. Due to particle-hole symmetry, the dispersion is odd
in $\mathbf{k}$. The zero-energy flat bands give a singular contribution to the
surface density of states, which can be detected as a sharp zero-bias peak in
the tunneling spectrum of an NCS--normal-metal
junction \cite{BST11,ScR11,STY11}. The shift of the surface bands in the TRSB
state causes a splitting of this peak, as shown in Fig.\ \ref{fig.disp}(b). This
splitting is a key experimental signature of TRSB. Indeed, the observed
splitting of the zero-bias peak for tunneling into the $(110)$ surface of the
cuprates is important evidence for TRSB in this system~\cite{Cov97,KD99}.

\begin{figure*}
\includegraphics[scale=0.2]{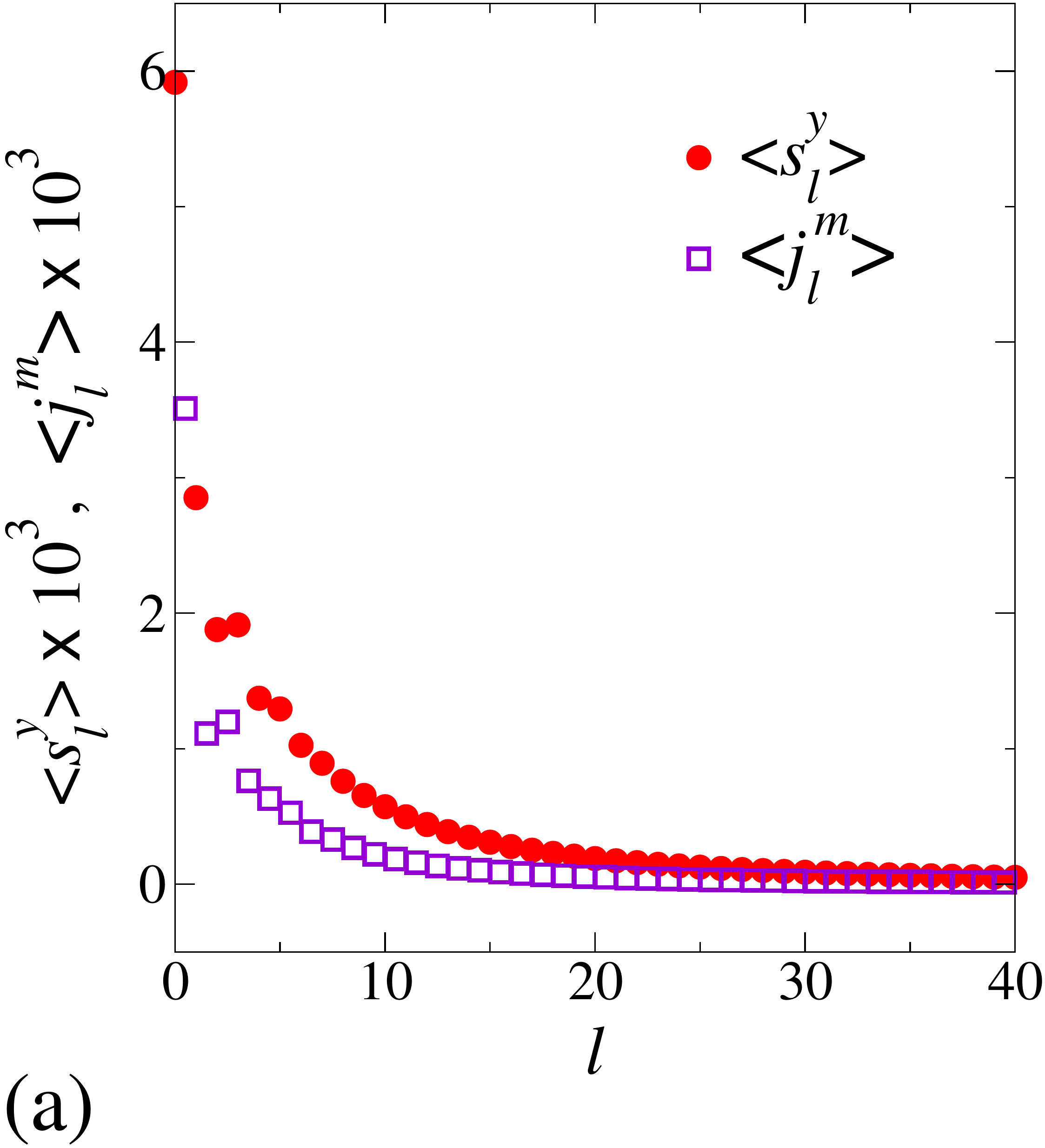}%
\hspace{1em}\includegraphics[width=0.73\columnwidth]{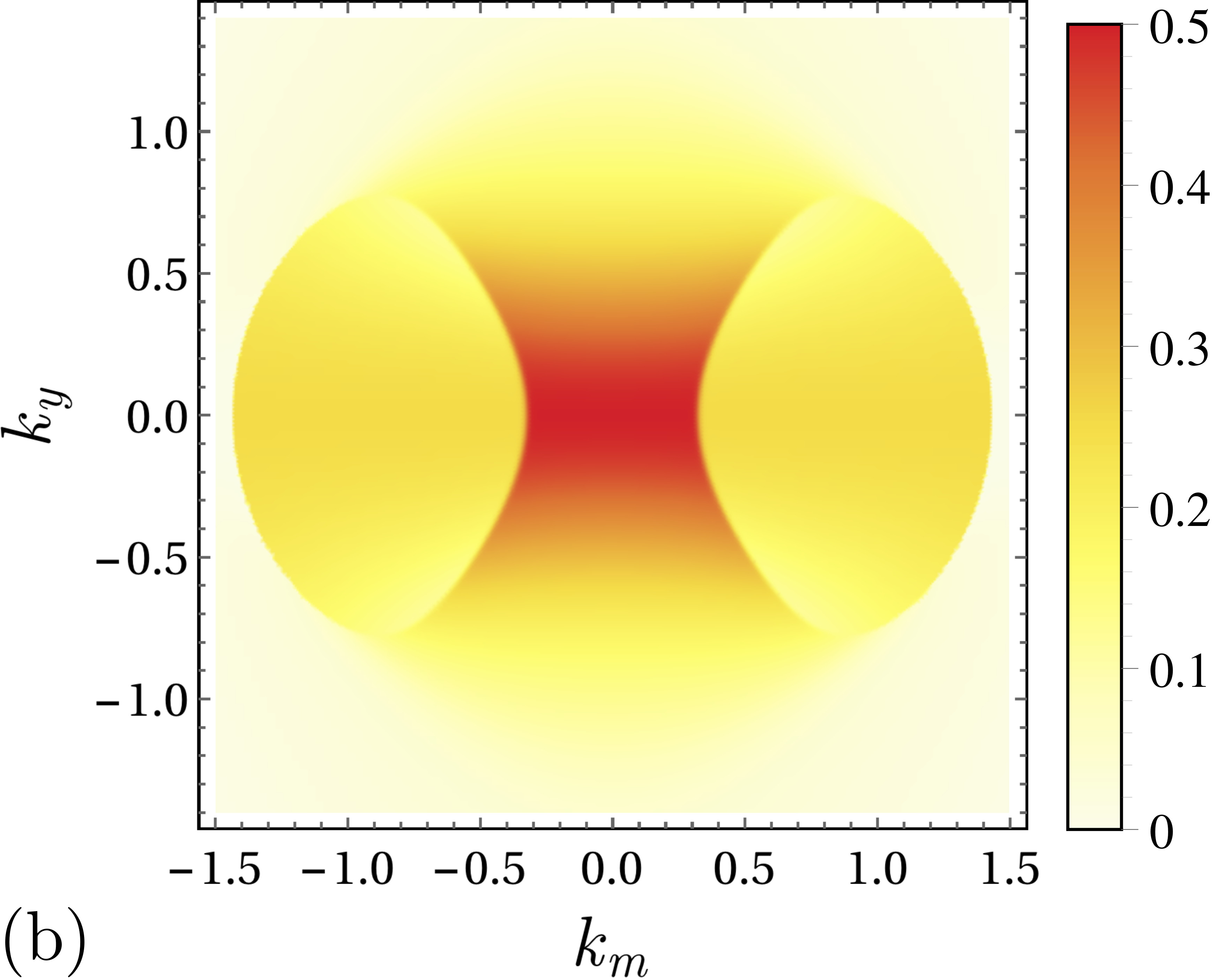}%
\hspace{1em}\includegraphics[width=0.73\columnwidth]{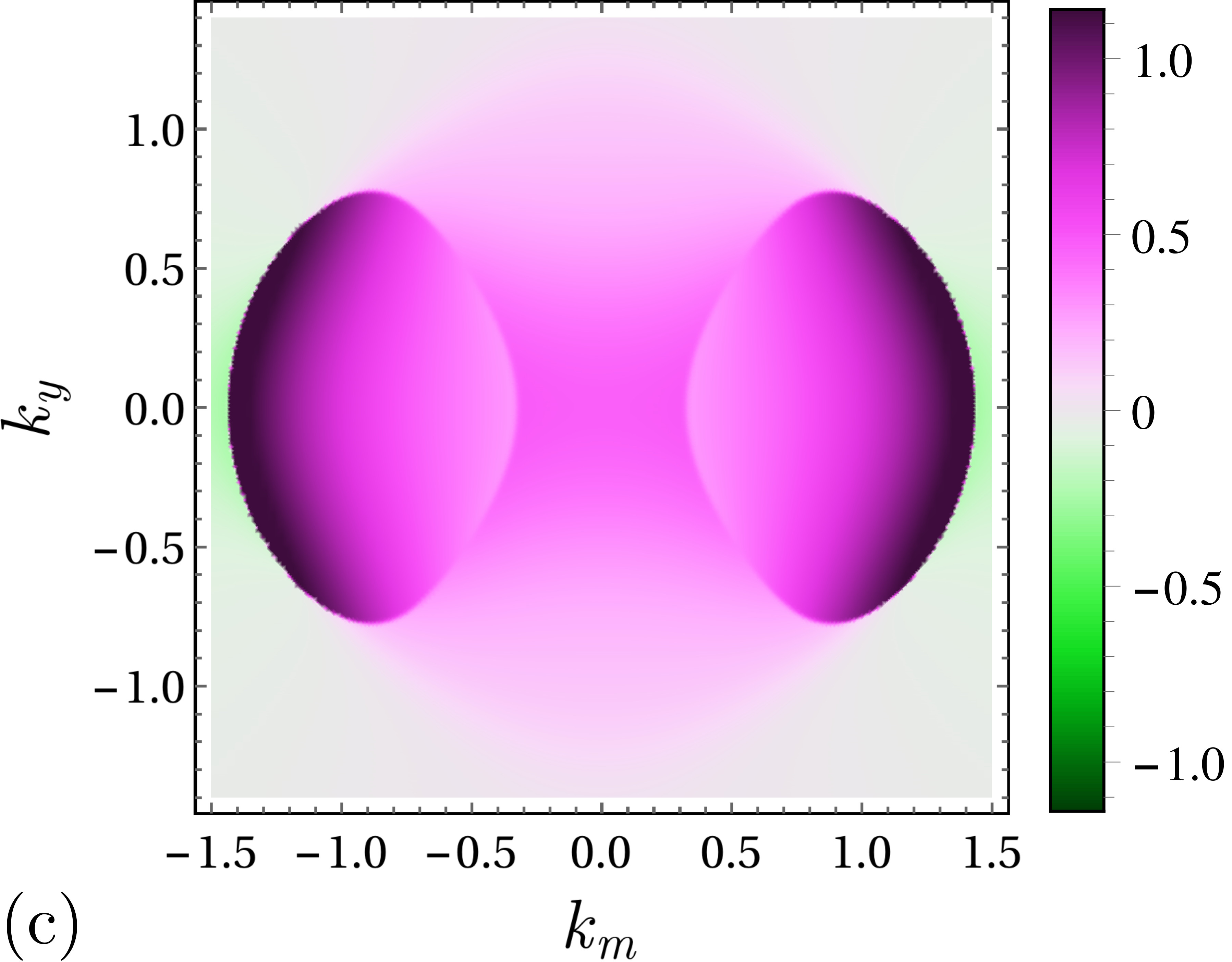}
\caption{(Color online) (a) spin polarization $\langle s^y_l\rangle$ (filled
circles) and current $\langle j^m_l\rangle$ (open squares) as functions of $l$,
for $W=300$ and the same parameters as in Fig.\ \ref{fig.Fermisurf}. Both
quantities are given in units of their value at the surface. The layer indices
of $\langle j^m_l\rangle$ are given as half integers to indicate that the
current flows between two layers, see the inset of Fig.\ \ref{fig.selfcgaps}.
(b) Momentum-resolved contributions to the $y$ component of the total spin
polarization of half the slab ($0\le l < W/2$) in the surface BZ. (c)
Momentum-resolved contributions to the $m$ component of the current in half the
slab ($0\le l < W/2$). The momentum-space plots in (b) and (c) are restricted to
a region just enclosing the projection of the positive-helicity Fermi surface.}
\label{fig.sycurm}
\end{figure*}


\textit{Spin polarization}.\ Broken TRS is also manifested by a nonzero spin
polarization near the surface, which is directed along the $y$-axis. A
polarization in other directions is forbidden by mirror symmetry in the $xz$
plane. Figure \ref{fig.sycurm}(a) shows the spatial variation of the
layer-resolved spin contributions $\langle s^y_l\rangle$; explicit expressions
for the spin operator $\mathbf{s}_l$ in layer $l$ and its thermal average are
given in Sec.\ {III} of the Supplemental Material \cite{suppl}. It is
interesting to examine how states at different $\mathbf{k}$ contribute to the
spin polarization: due to the strong polarization of the flat-band surface
states in the TRS state \cite{QuS14,BST15}, one might expect that the spin
polarization largely originates from the shifted flat bands. To check this, we
plot in Fig.\ \ref{fig.sycurm}(b) the momentum-resolved contribution to the spin
polarization of the half slab defined by $0\le l< W/2$ \cite{suppl}.
Surprisingly, the spin polarization is not primarily carried by the shifted flat
bands but rather by bulk and perhaps dispersing surface states
\cite{BST11,SBT12} from the region between the projected nodal rings.

\textit{Equilibrium currents}.\ Furthermore, the absence of TRS permits a
nonzero equilibrium surface current \cite{MaS95,Sig98,HWS00}. Indeed, we expect
such a current since the surface bands become dispersive and the dispersion is
odd in $k_m$; a similar modification of the electronic structure at an interface
with a ferromagnet does result in a surface current \cite{BTS13,STB13}. Explicit
expressions for the current operator $\mathbf{j}_l$ in layer $l$ and its thermal
average are given in Sec.\ {IV} of the Supplemental Material \cite{suppl}.
Although charge is not conserved in the superconducting MF state, one can
account for the pairing potentials by adding so-called source terms to the
continuity equation~\cite{FuT91}. For self-consistently calculated gaps,
however, the thermal average of the source terms vanishes, and charge
conservation is retained \cite{FuT91}. This implies that the current
perpendicular to the slab's surface, i.e., in the $l$ direction, must vanish.
Mirror symmetry in the $xz$ plane forbids a current along the $y$-axis
\cite{suppl}, leaving only the current along the $m$ direction, defined as
$\langle j^m_{l+1/2}\rangle = (\langle j^x_{l+1/2}\rangle
  - \langle j^z_{l+1/2}\rangle)/\sqrt{2}$.
$\langle j^m_{l+1/2} \rangle$ is indeed nonzero in the TRSB state: In Fig.\
\ref{fig.sycurm}(a) we plot the current as a function of the layer index $l$,
which shows that it is bound to the surface with spatial profile similar to the
spin polarization. In contrast to the spin polarization, the main
contribution to the current stems from surface states within the projected nodal
rings, as shown by the momentum-resolved current in a half slab plotted in
Fig.\ \ref{fig.sycurm}(c). We have also studied the contributions to the
vanishing components $\langle j^l_l\rangle$ and $\langle j^y_l\rangle$, shown in
the Supplemental Material~\cite{suppl}.
Interestingly, $\langle j^l_l\rangle$ cancels only in the sum over the
full surface BZ, showing that bulk states must be included to satisfy charge
conservation. Note that the sign of both the spin polarization and the current
is reversed for the degenerate solution with complex-conjugated gaps.

The coupling to the electromagnetic field, which is not included here, leads to
additional screening currents  that exactly balance the spontaneous surface
current in the limit $W\to\infty$. However, these currents build up on the
length scale of the magnetic penetration depth $\lambda$, which in typical NCSs
is much larger than the decay length of the surface current, on the order of the
coherence length $\xi$ \cite{ncsbook}. In samples with thickness smaller than
the penetration depth but larger than the coherence length, it should thus be
possible to detect the surface current.

\textit{Summary and conclusions}.\ We have studied the stability of zero-energy
flat bands at the surface of an NCS within selfconsistent MF theory. We find
that the flat bands are indeed recovered by the selfconsistent calculation
within a broad temperature range below the bulk transition temperature $T_c$.
TRS is spontaneously broken at a much lower temperature $T_s$, which is signaled
by a nonuniform phase of the gaps. This destroys the topological protection for
the flat bands, shifting them away from zero energy and giving them finite
velocity. Figure \ref{fig.disp} shows that at low temperatures the flat bands
are displaced by an energy on the order of $T_s$, which is significantly smaller
than the bulk gaps of order $T_c$. The free energy gain due to the shift of the
flat bands is likely a major driver of the TRSB state, and ultimately limits
$T_s$ as the free energy gain from the shift is reduced by the broadening of the
Fermi function.

The TRSB state leads to clear experimental signatures: a splitting of the
zero-bias peak in the tunneling spectrum, a nonvanishing spin polarization at
the surface, and a nonvanishing equilibrium charge current parallel to the
surface. The latter two effects show that the TRSB state found here is
qualitatively different from that predicted for the $(110)$ surface of cuprate
superconductors~\cite{MaS95,FRS97,ZFT99,HWS00,Sig98,PL14}.

\textit{Acknowledgements}.\ We thank A. P. Schnyder, R. Quieroz, and T. Neupert
for helpful discussions. C. T. gratefully aknowledges support by the Deut\-sche
For\-schungs\-ge\-mein\-schaft through Research Training Group GRK 1621 and
Collaborative Research Center SFB 1143. S. R. was supported by the Norwegian
Research Council, Grant Nos.\ 205591/V20 and 216700/F20. P. M. R. B.
acknowledges support from Microsoft Station Q, LPS-CMTC, and JQI-NSF-PFC.

\onecolumngrid


\newpage
\widetext

\renewcommand{\theequation}{S\arabic{equation}}
\renewcommand{\thefigure}{S\arabic{figure}}
\setcounter{page}{1}
\setcounter{equation}{0}
\setcounter{figure}{0}

\begin{center}
\textbf{{\large Supplemental Material for\\[0.5ex]
Surface instability in nodal noncentrosymmetric superconductors}}\\[1.5ex]
Carsten Timm, Stefan Rex, and P. M. R. Brydon
\end{center}

\section{Mean-field theory for the bulk}

In this section we sketch the MF theory for the bulk NCS. We assume spatially
uniform pairing potentials $\Delta^s_j = \Delta_s$ and $\Delta^t_{ij} =
\Delta_t$. Using this ansatz to decouple the interaction Hamiltonian
$H_\mathrm{int}$, we obtain the Bogoliubov-de Gennes (BdG)
Hamiltonian~\cite{S.SBT12}
\begin{equation}
H_\mathrm{MF} = \frac12\, \sum_\mathbf{k} \Phi^\dagger_\mathbf{k}
  \mathcal{H}(\mathbf{k}) \Phi_\mathbf{k}
  + N\, \frac{\Delta_s^2}{U_s} + N\, \frac{\Delta_t^2}{U_t} ,
\end{equation}
with the number of sites, $N$, and the block matrix
\begin{equation}
\mathcal{H}(\mathbf{k}) = \left(\begin{array}{cc}
  h(\mathbf{k}) & \Delta(\mathbf{k}) \\
  \Delta^\dagger(\mathbf{k}) & -h^T(-\mathbf{k})
  \end{array}\right) \label{eq:Hamk}
\end{equation}
written in terms of
$h(\mathbf{k}) = \xi_\mathbf{k} \sigma^0 - \lambda\,
  \mathbf{l}_\mathbf{k}\cdot \mbox{\boldmath$\sigma$}$,
$\Delta(\mathbf{k}) = (\Delta_s \sigma^0 + \Delta_t\, \mathbf{l}_\mathbf{k}
  \cdot \mbox{\boldmath$\sigma$})\, i\sigma^y$,
$\xi_{\mathbf{k}} = -2t\,(\cos k_x + \cos k_y + \cos k_z) - \mu$,
$\mathbf{l}_\mathbf{k} = \hat\mathbf{x}\,\sin k_y - \hat\mathbf{y}\, \sin k_x$,
and the Nambu spinor
$\Phi_\mathbf{k} = (c_{\mathbf{k}\uparrow},c_{\mathbf{k}\downarrow},
  c^\dagger_{-\mathbf{k},\uparrow},c^\dagger_{-\mathbf{k},\downarrow})^T$.
Here, $\sigma^0$ is the $2\times 2$ identity matrix. The dispersion
$E_{\mathbf{k}\nu}$, $\nu=1,\ldots,4$ is obtained by diagonalizing
$\mathcal{H}(\mathbf{k})$. $\Delta_s$ and $\Delta_t$ are then obtained by
minimizing the free energy
\begin{equation}
F_\mathrm{MF} = -k_BT \left.\sum_{\mathbf{k}\nu}\right.'
  \ln \left( 2 \cosh \frac{\beta E_{\mathbf{k}\nu}}{2} \right)
  + N\, \frac{\Delta_s^2}{U_s} + N\, \frac{\Delta_t^2}{U_t} ,
\end{equation}
where the momentum sum is over half the BZ, $k_m>0$. This restriction of the sum
makes use of particle-hole symmetry, which relates the Hamiltonian in Eq.\
(\ref{eq:Hamk}) at $\mathbf{k}$ and $-\mathbf{k}$ by \cite{S.SBT12}
$\mathcal{U}_C\,\mathcal{H}^T(-\mathbf{k})\, \mathcal{U}_C^\dagger =
-\mathcal{H}(\mathbf{k})$
with the unitary matrix $\mathcal{U}_C = \sigma^x \otimes \sigma^0$.

\section{Mean-field theory for the slab}

We now set up the MF Hamiltonian for the (101) slab and describe the
determination of the gap parameters $\Delta^s_l$, $\Delta^x_{l+1/2}$, and
$\Delta^y_l$ in the MF approximation. After Fourier transformation in the
directions parallel to the surfaces, the MF Hamiltonian reads
\begin{eqnarray}
H_\mathrm{MF} &=& \frac12 \sum_{\mathbf{k}} \sum_{l=0}^{W-1}
  \Phi^\dagger_{\mathbf{k}l} \mathcal{H}_{ll}(\mathbf{k}) \Phi_{\mathbf{k}l}
  + \frac12 \sum_{\mathbf{k}} \sum_{l=0}^{W-2}
    \Phi^\dagger_{\mathbf{k},l+1} \mathcal{H}_{l+1,l}(\mathbf{k})
    \Phi_{\mathbf{k}l}
  + \frac12 \sum_{\mathbf{k}} \sum_{l=1}^{W-1}
    \Phi^\dagger_{\mathbf{k},l-1} \mathcal{H}_{l-1,l}(\mathbf{k})
    \Phi_{\mathbf{k}l} \nonumber \\
&& {}+ \frac{N_\|}{U_s} \sum_{l=0}^{W-1} |\Delta^s_l|^2
  + \frac{N_\|}{2U_t} \sum_{l=0}^{W-2} |\Delta^x_{l+1/2}|^2
  + \frac{N_\|}{2U_t} \sum_{l=0}^{W-1} |\Delta^y_l|^2 ,
\end{eqnarray}
where $N_\|$ is the number of unit cells of the slab and $\Phi_{\mathbf{k}l} =
(c_{\mathbf{k}l\uparrow},c_{\mathbf{k}l\downarrow},
c^\dagger_{-\mathbf{k},l,\uparrow},c^\dagger_{-\mathbf{k},l,\downarrow})^T$ is
the partially Fourier-transformed Nambu spinor. The sums over $l$ containing
$\Phi^\dagger_{\mathbf{k},l\pm 1}$ are restricted in such a way that $l\pm 1\in
\{0,\ldots,W-1\}$. The coefficient matrices appearing in $H_\mathrm{MF}$ are
\begin{eqnarray}
\mathcal{H}_{ll}(\mathbf{k}) &=& \left(\begin{array}{cccc}
  -2t \cos k_y - \mu & -\lambda\sin k_y & -\Delta^y_l \sin k_y & \Delta^s_l \\
  -\lambda\sin k_y & -2t \cos k_y - \mu & -\Delta^s_l & \Delta^y_l \sin k_y \\
  -\Delta^{y*}_l \sin k_y & -\Delta^{s*}_l & 2t\cos k_y+\mu & -\lambda\sin k_y\\
  \Delta^{s*}_l & \Delta^{y*}_l \sin k_y & -\lambda\sin k_y & 2t\cos k_y+\mu
  \end{array}\right) ,
\label{Hk.3a} \\
\mathcal{H}_{l\pm 1,l}(\mathbf{k}) &=& \left(\begin{array}{cccc}
  -2t \cos(k_m/\sqrt{2}) & \pm (\lambda/2)\, e^{\mp ik_m/\sqrt{2}} &
    \pm (\Delta^x_{l\pm 1/2}/2)\, e^{\mp ik_m/\sqrt{2}} & 0 \\
  \mp (\lambda/2)\, e^{\mp ik_m/\sqrt{2}} & -2t \cos(k_m/\sqrt{2}) &
    0 & \pm (\Delta^x_{l\pm 1/2}/2)\, e^{\mp ik_m/\sqrt{2}} \\
  \mp (\Delta^{x*}_{l\pm 1/2}/2)\, e^{\mp ik_m/\sqrt{2}} & 0 &
    2t \cos(k_m/\sqrt{2}) & \mp (\lambda/2)\, e^{\mp ik_m/\sqrt{2}} \\
  0 & \mp (\Delta^{x*}_{l\pm 1/2}/2)\, e^{\mp ik_m/\sqrt{2}} &
    \pm (\lambda/2)\, e^{\mp ik_m/\sqrt{2}} & 2t \cos(k_m/\sqrt{2})
  \end{array}\right) .\qquad
\label{Hk.3b}
\end{eqnarray}
We next construct the $4W\times 4W$ block matrix
\begin{equation}
\mathcal{H}(\mathbf{k}) \equiv \left( \begin{array}{cccc}
  \mathcal{H}_{00}(\mathbf{k}) & \mathcal{H}_{01}(\mathbf{k}) & 0 & \cdots \\
  \mathcal{H}_{10}(\mathbf{k}) & \mathcal{H}_{11}(\mathbf{k}) &
    \mathcal{H}_{12}(\mathbf{k}) & \cdots \\
  0 & \mathcal{H}_{21}(\mathbf{k}) & \mathcal{H}_{22}(\mathbf{k}) & \cdots \\
  \vdots & \vdots & \vdots & \ddots
  \end{array}\right)
\label{Hblockk.2}
\end{equation}
and denote its eigenvalues by $E_{\mathbf{k}\nu}$, $\nu=1,\ldots,4W$ and the
corresponding eigenvectors by $|\mathbf{k}\nu\rangle$.
The MF Hamiltonian satisfies particle-hole symmetry \cite{S.SBT12},
$\mathcal{U}_C\, \mathcal{H}^T(-\mathbf{k})\, \mathcal{U}_C^\dagger =
  - \mathcal{H}(\mathbf{k})$
with the unitary matrix $\mathcal{U}_C = \mathbbm{1}_W \otimes \sigma^x \otimes
\sigma^0$, where $\mathbbm{1}_W$ is the $W\times W$ identity matrix. This
symmetry again allows to restrict the momentum sums to half the BZ. The free
energy can then be written as
\begin{equation}
F_\mathrm{MF} = -k_BT \left.\sum_{\mathbf{k}\nu}\right.'
  \ln \left( 2 \cosh \frac{\beta E_{\mathbf{k}\nu}}{2} \right)
  + \frac{N_\|}{U_s} \sum_{l=0}^{W-1} |\Delta^s_l|^2
  + \frac{N_\|}{2U_t} \sum_{l=0}^{W-2} |\Delta^x_{l+1/2}|^2
  + \frac{N_\|}{2U_t} \sum_{l=0}^{W-1} |\Delta^y_l|^2 ,
\label{FMF.3}
\end{equation}
where the momentum sum is restricted to half the BZ, $k_m>0$. Minimization of
$F_\mathrm{MF}$ gives the gaps $\Delta^s_l$, $\Delta^x_{l+1/2}$, and
$\Delta^y_l$. The derivatives of $F_\mathrm{MF}$ with respect to the complex
conjugate gaps can be calculated with the help of the Hellmann-Feynman theorem,
for example
\begin{equation}
\frac{\partial F_\mathrm{MF}}{\partial\Delta^{s*}_l}
  = -\frac{1}{2} \left.\sum_{\mathbf{k}\nu}\right.'
  \tanh \frac{\beta E_{\mathbf{k}\nu}}{2} \,
  \langle \mathbf{k}\nu| \,
  \frac{\partial\mathcal{H}(\mathbf{k})}{\partial\Delta^{s*}_l}
  |\mathbf{k}\nu\rangle
  + \frac{N_\|}{U_s}\, \Delta^s_l .
\end{equation}
The momentum sums are performed on a $50\times50$ mesh, referring to the full
surface BZ. Quadrupling the number of points in the mesh to $100\times 100$
leads to changes in the MF gaps on the order of only $0.1$\%.

Solving the resulting MF equations by iteration turns out to be prohibitively
slow for the required $W$, essentially because the minimum of $F_\mathrm{MF}$ is
very shallow in some directions in the high-dimensional space of gap parameters.
On the other hand, numerical minimization making use of the explicitly known
gradient is reasonably efficient. We use the Broyden-Fletcher-Goldfarb-Shanno
method implemented in Numerical Recipes \cite{S.NR}. It requires an initial
guess for the inverse Hessian. When we scan over ranges of temperatures, we use
not only the converged values of the gaps but also the best approximate inverse
Hessian from one step as starting values for the next, which significantly
speeds up the convergence. We assume that the method has converged when no real
or imaginary part of any gap parameter changes by more than (double) machine
precision in the last step.

For certain parameter values, we find nonvanishing gradients of the phases of
the order parameters in the $l$ direction, normal to the surfaces.
Specifically, we find four metastable solutions, which are mapped onto each
other by inverting the phase gradients at one or both surfaces. In the limit
$W\to\infty$, the four solutions are degenerate. For finite $W$, they split into
two degenerate pairs with phase gradients that are even and odd, respectively,
under reflection at the center of the slab. We here choose a solution with even
phase gradients since then the selfconsistent solution ensures that the phases
of $\Delta^s_l$, $\Delta^x_{l+1/2}$, and $\Delta^y_l$ become equal at the center
of the slab; equal phases of all gaps at the center are expected since the bulk
MF solution has equal phases. By a global phase change we can then make the
phase of all gaps zero at the center. The phases and imaginary parts of the gaps
are then odd under reflection at the center. Finally, of the two remaining
solutions differing in the sign of the imaginary parts of the gaps, we select
the solution with $\mathrm{Im}\,\Delta^s_0\ge 0$ for definiteness. The other
solution leads to inverted spin polarizations and currents.

\section{Spin polarization}

Here, we present expressions for the spin polarization. The operator of the spin
per site, averaged over the directions parallel to the surfaces, is
\begin{equation}
\mathbf{s}_l = \frac{1}{N_\|} \sum_\mathbf{k}
  c^\dagger_{\mathbf{k}l}\, \frac{\mbox{\boldmath$\sigma$}}{2}\,
  c_{\mathbf{k}l} .
\end{equation}
Using particle-hole symmetry, the thermal spin average can be written as
\begin{equation}
\langle\mathbf{s}_l\rangle = -\frac{1}{4N_\|}
  \left.\sum_{\mathbf{k}\nu}\right.'
  \tanh\frac{\beta E_{\mathbf{k}\nu}}{2} \,
  \langle \mathbf{k}\nu|
  P_{ll} \otimes \left(\begin{array}{cc}
    \mbox{\boldmath$\sigma$} & 0 \\
    0 & -\mbox{\boldmath$\sigma$}^T\!
  \end{array}\right) |\mathbf{k}\nu\rangle ,
\label{sofk.3}
\end{equation}
where $P_{ll'}$ is a $W\times W$ matrix with the components $(P_{ll'})_{nn'} =
\delta_{ln} \delta_{l'n'}$. We also consider the momentum-dependent
contributions to the spin polarization of the half slab defined by $0\le l<
W/2$. These contributions are obtained by summing $\langle \mathbf{s}_l\rangle$
over $l=0,\ldots,W/2-1$ and removing the factor $1/N_\|$ and the momentum sum.

\section{Equilibrium current}

The second observable of interest is the current. The operators $j^\alpha_{ij}$
denote the \emph{electron-number} current from site $j$ to its nearest neighbor
$i$ in the $\alpha=x,y,z$ direction. They can be read off from $H_0$ in Eq.\
({1}) in the main text,
%
%
\begin{eqnarray}
j^x_{ij} &=& -i\, c^\dagger_i \left(\begin{array}{cc}
  -t & \lambda/2 \\ -\lambda/2 & -t
  \end{array}\right) c_j
  + i\, c^\dagger_j \left(\begin{array}{cc}
  -t & -\lambda/2 \\ \lambda/2 & -t
  \end{array}\right) c_i , \\
j^y_{ij} &=& -i\, c^\dagger_i \left(\begin{array}{cc}
  -t & -i\lambda/2 \\ -i\lambda/2 & -t
  \end{array}\right) c_j
  + i\, c^\dagger_j \left(\begin{array}{cc}
  -t & i\lambda/2 \\ i\lambda/2 & -t
  \end{array}\right) c_i , \\
j^z_{ij} &=& -i\, c^\dagger_i \left(\begin{array}{cc}
  -t & 0 \\ 0 & -t
  \end{array}\right) c_j
  + i\, c^\dagger_j \left(\begin{array}{cc}
  -t & 0 \\ 0 & -t
  \end{array}\right) c_i .
\end{eqnarray}
The interaction term $H_\mathrm{int}$ conserves charge locally and therefore
does not contribute to the current operator. After the MF decoupling, the
anomalous terms do not conserve charge---they describe creation or annihilation
of two electrons either at the same site or at neighboring sites. Such processes
do not lead to currents but do introduce a source term, which is discussed in
the main text. We average the current over layers parallel to the surface,
taking into account that $j^x_{ij}$ and $j^z_{ij}$ connect adjacent layers,
whereas $j^y_{ij}$ describes a current within a single layer. We then obtain the
thermal averages, again using particle-hole symmetry,
\begin{eqnarray}
\langle j^x_{l+1/2} \rangle &=& -\frac{1}{2N_\|}
  \left.\sum_{\mathbf{k}\nu}\right.'
  \tanh\frac{\beta E_{\mathbf{k}\nu}}{2} \,
  \langle \mathbf{k}\nu| \left\{ i\, e^{-ik_m/\sqrt{2}}\,
  P_{l+1,l} \otimes \left(\begin{array}{cccc}
    t & -\lambda/2 & 0 & 0 \\
    \lambda/2 & t & 0 & 0 \\
    0 & 0 & t & -\lambda/2 \\
    0 & 0 & \lambda/2 & t
    \end{array}\right) + \mathrm{H.c.} \right\}  |\mathbf{k}\nu\rangle , \\
\langle j^y_l \rangle &=& -\frac{1}{N_\|}
  \left.\sum_{\mathbf{k}\nu}\right.'
  \tanh\frac{\beta E_{\mathbf{k}\nu}}{2} \,
  \langle \mathbf{k}\nu|\, P_{ll} \otimes \left(\begin{array}{cccc}
    t \sin k_y & -(\lambda/2) \cos k_y & 0 & 0 \\
    -(\lambda/2) \cos k_y & t \sin k_y & 0 & 0 \\
    0 & 0 & t \sin k_y & (\lambda/2) \cos k_y \\
    0 & 0 & (\lambda/2) \cos k_y & t \sin k_y
    \end{array}\right)  |\mathbf{k}\nu\rangle ,\qquad \\
\langle j^z_{l+1/2} \rangle &=& -\frac{1}{2N_\|}
  \left.\sum_{\mathbf{k}\nu}\right.'
  \tanh\frac{\beta E_{\mathbf{k}\nu}}{2} \,
  \langle \mathbf{k}\nu| \left\{ i\, e^{ik_m/\sqrt{2}}\,
  P_{l+1,l} \otimes \left(\begin{array}{cccc}
    t & 0 & 0 & 0 \\
    0 & t & 0 & 0 \\
    0 & 0 & t & 0 \\
    0 & 0 & 0 & t
    \end{array}\right) + \mathrm{H.c.} \right\}  |\mathbf{k}\nu\rangle ,
\end{eqnarray}
where $\langle j^{x,z}_{l+1/2}\rangle$ denotes currents connecting layers $l$
and $l+1$. The components with respect to the slab coordinates are
\begin{equation}
\langle j^l_{l+1/2}\rangle = \frac{\langle j^x_{l+1/2}\rangle + \langle
  j^z_{l+1/2}\rangle}{\sqrt{2}} , \qquad
\langle j^m_{l+1/2}\rangle = \frac{\langle j^x_{l+1/2}\rangle - \langle
  j^z_{l+1/2}\rangle}{\sqrt{2}} .
\end{equation}
We note that $\langle j^y_l\rangle$ vanishes for any choice of gap parameters
for our model, even non-selfconsistent ones. This is based on mirror symmetry in
the $xz$ plane. The current in the $y$ direction changes sign under this
symmetry operation and thus vanishes.

\begin{figure}
\raisebox{1ex}{(a)}\includegraphics[width=2.7in]{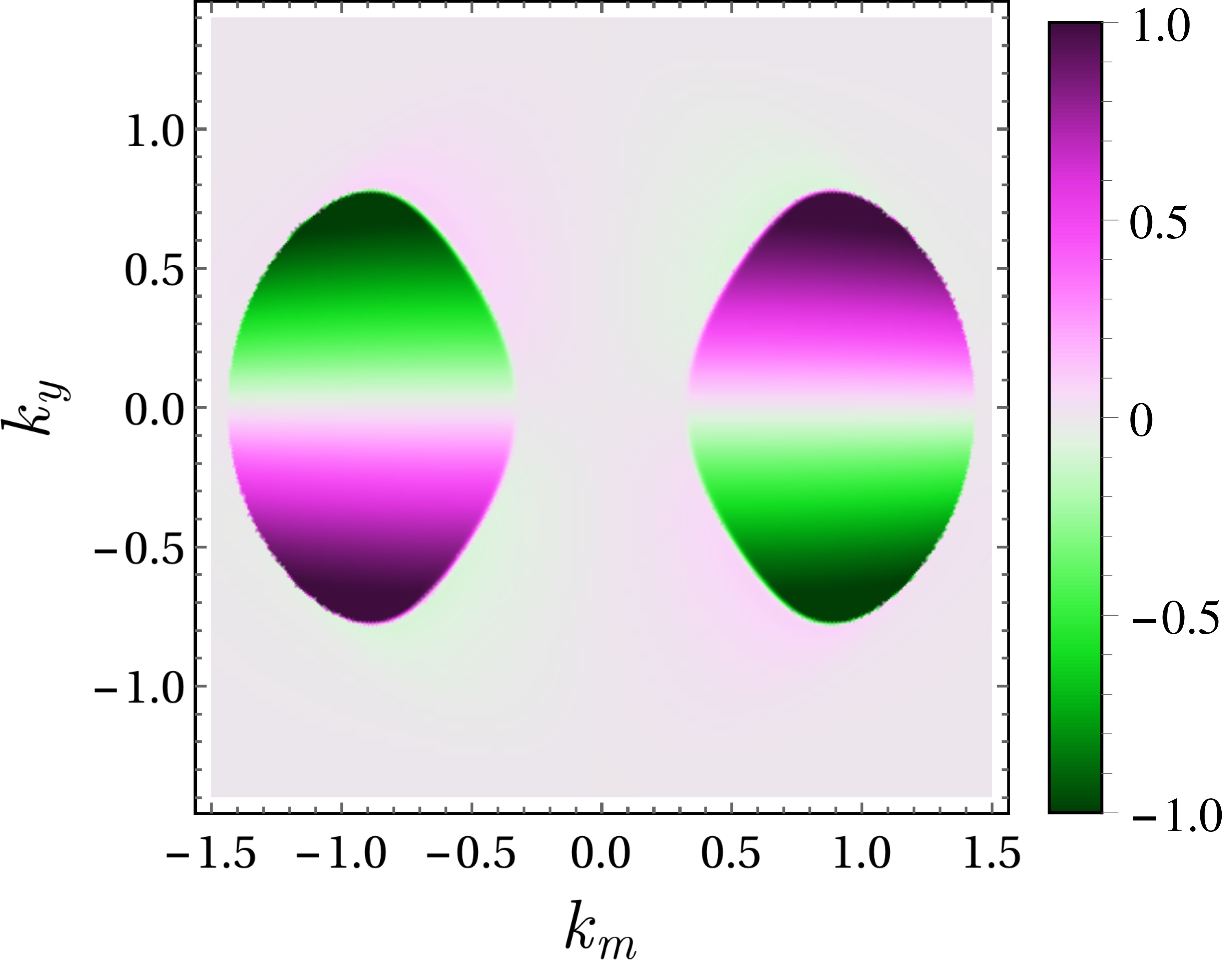}%
\hspace{1em}\raisebox{1ex}{(b)}\includegraphics[width=2.7in]{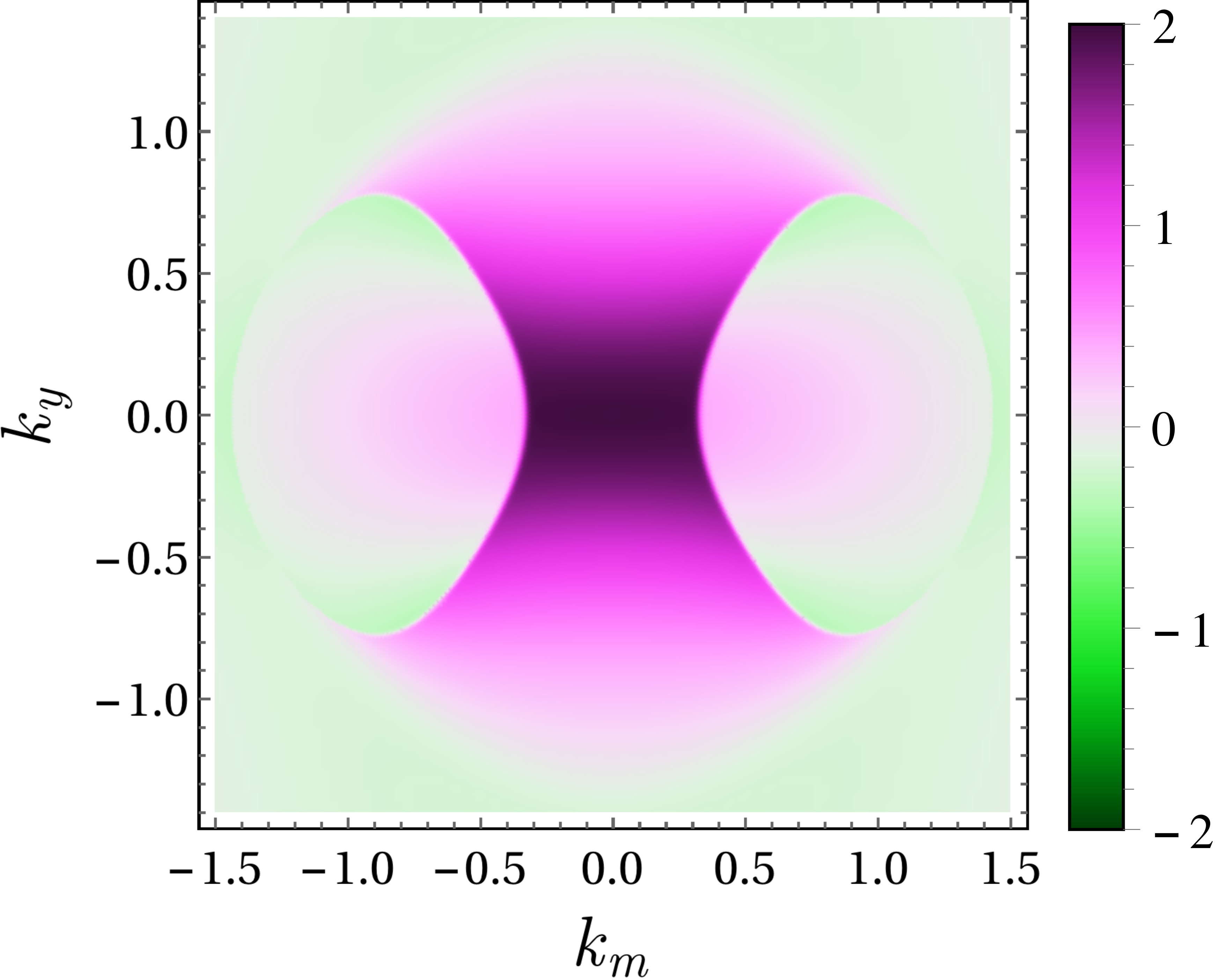}
\caption{Momentum-resolved contributions to (a) the $y$ component and (b) the
$l$ component of the current in half the slab ($0\le l < W/2$).}
\label{fig.curyk.curlk}
\end{figure}

The momentum-dependent contributions to the current in the half slab $0\le l<
W/2$ are obtained by summing $\langle \mathbf{j}_l\rangle$ over
$l=0,\ldots,W/2-1$ and removing the factor $1/N_\|$ and the momentum sum. The
momentum-resolved $m$
component, which sums to a nonzero current, is shown in Fig.\ {5}(c) in the main
text. We present the momentum-resolved $y$ and $l$ components in Fig.\
\ref{fig.curyk.curlk}. The $y$ components chancel by symmetry, as noted above.
The cancelation of the $l$ components, which is required by charge conservation,
is only ensured for selfconsistent gaps \cite{S.FuT91}. Large positive
contributions from bulk states within the projected (small) positive-helicity
Fermi surface are canceled by small negative contributions from the flat bands
and from bulk states within the projected (large) negative-helicity Fermi
surface. This shows that the bulk states must be included to satisfy charge
conservation.
%

\begin{references}
\bibitem{endnote.lambda}Note that the sign of the  SOC is inverted compared to
Refs.~\cite{BST11,SBT12,BST15}.
\end{references}
\end{document}